\documentclass[conference]{IEEEtran}
\IEEEoverridecommandlockouts
\usepackage{cite}
\usepackage{makecell}
\usepackage{amsmath,amssymb,amsfonts}
\usepackage{algorithmic}
\usepackage{graphicx}
\usepackage{textcomp}
\usepackage{xcolor}
\def\BibTeX{{\rm B\kern-.05em{\sc i\kern-.025em b}\kern-.08em
    T\kern-.1667em\lower.7ex\hbox{E}\kern-.125emX}}

\usepackage[utf8]{inputenc}
\usepackage{times}
\usepackage{soul}
\usepackage{url}
\usepackage[hidelinks]{hyperref}
\usepackage[utf8]{inputenc}
\usepackage[small]{caption}
\usepackage{booktabs}
\usepackage{algorithm}
\usepackage{enumitem}

\usepackage[absolute,overlay]{textpos}
\usepackage{amsthm}

\newtheorem{remark}{Remark}
\newtheorem{example}{Example}

\newtheorem{definition}{Definition}

\newcount\Comments  
\usepackage{color}
\definecolor{darkgreen}{rgb}{0,0.5,0}
\definecolor{purple}{rgb}{1,0,1}
\newcommand{\kibitz}[2]{\ifnum\Comments=1\textcolor{#1}{#2}\fi}

\newcommand{\robab}[1]{\kibitz{blue}       {[robab: #1]}}

\usepackage{pdflscape}
\usepackage{comment}

\usepackage{multirow}
\usepackage{tabularx}

\begin{document}


\title{On the Need for a Statistical Foundation in Scenario-Based Testing of Autonomous Vehicles
\thanks{Supported by the UK EPSRC New Investigator Award [EP/Z536568/1].
}
}

\author{
\IEEEauthorblockN{
Xingyu~Zhao$^{1}$, Robab~Aghazadeh-Chakherlou$^1$, 
Chih-Hong Cheng$^2$, 
Peter Popov$^3$, 
Lorenzo~Strigini$^3$}
  \IEEEauthorblockA{
  \textit{$^1$WMG, University of Warwick, 
  Coventry, U.K.} \\
  \textit{$^2$Department of Computer Science, Chalmers University of Technology \& University of Gothenburg, Gothenburg, Sweden}\\
  \textit{$^3$Centre for Software Reliability, City St George's, University of London, London, U.K.} \\
  \{xingyu.zhao,robab.aghazadeh-chakherlou\}@warwick.ac.uk, chihhong@chalmers.se, \{p.t.popov,l.strigini\}@city.ac.uk}

}


\maketitle

 \begin{textblock*}{20cm}(1cm,1cm) \textcolor{red}{Accepted by ITSC 2025: 
 \url{https://ieee-itsc.org/2025/}}
 \end{textblock*}

\begin{abstract}
Scenario-based testing has emerged as a common method for autonomous vehicles (AVs) safety assessment, offering a more efficient alternative to mile-based testing by focusing on high-risk scenarios. However, fundamental questions persist regarding its stopping rules, residual risk estimation, debug effectiveness, and the impact of simulation fidelity on safety claims. This paper argues that a rigorous statistical foundation is essential to address these challenges and enable rigorous safety assurance. By drawing parallels between AV testing and established software testing methods, we identify shared research gaps and reusable solutions. We propose proof-of-concept models to quantify the probability of failure per scenario (\textit{pfs}) and evaluate testing effectiveness under varying conditions. Our analysis reveals that neither scenario-based nor mile-based testing universally outperforms the other. Furthermore, we give an example of formal reasoning about alignment of synthetic and real-world testing outcomes, a first step towards supporting statistically defensible simulation-based safety claims. 

\end{abstract}


\section{Introduction}

Autonomous vehicles (AVs) are transitioning rapidly from research environments to public roads. Waymo initiated this shift by launching its first commercial AV taxi in December 2018. More recently, Tesla has announced plans to deploy a robotaxi service in Austin, Texas, by June 2025, using its Full Self-Driving system. These developments underscore the urgent need for comprehensive and rigorous safety assessment to satisfy regulatory standards and public safety expectations. 

While many efforts have been made to address the AV safety challenges covering various aspects like design, implementation, regulation and legal issues~\cite{anderson_autonomous_2016,fagnant_preparing_2015,koopman_autonomous_2017,bonnefon_social_2016,wang_survey_2024}, AV testing remains an indispensable method for generating the verification and validation evidence necessary for safety assurance. In general, there are two common types of AV testing, \textit{mile-based} and \textit{scenario-based}\footnote{Scenario-based testing may be executed by conducting the tests on different ``platforms'': simulators, ``proving grounds'' or on the public roads.}. The former involves accumulating a large number of real-world or simulated miles to statistically validate the vehicle’s safety. The premise is that, as the vehicle encounters a wider variety of driving conditions and edge cases, statistical confidence in its safety increases. Given the required rarity of incidents, demonstrating safety through this method would require millions/billions of miles~\cite{kalra_driving_2016,zhao2019assessing,littlewood_validation_1993}. The industry largely resists such mile-based ``driving to safety'' strategy due to the prohibitive cost and risk. Instead, they favour
scenario-based testing~\cite{ulbrich2015defining,iso_scenario_2022,bsi_scenario_2023,Riedmaier_2020} as a complementary approach, which focuses on using controlled, high-risk scenarios to infer the AV's safety. This method avoids the need for extensive driven miles; instead, it tests the vehicle’s response to predefined, potentially dangerous situations through simulated or field-tested extreme scenarios.

However, current scenario-based testing practices cannot yet answer several pressing questions:

\textit{i)} \textit{What is the stopping rule of scenario-based testing?} The number of possible parameter combinations to describe test scenarios is exploding, so exhaustive testing of all possible scenarios for a given operational design domain (ODD) is infeasible~\cite{amersbach2019defining}. While coverage criteria were studied (e.g. in~\cite{amersbach2019defining,9239916,8317919}) for scenario-based testing, borrowing ideas from traditional software testing where coverage criteria are used to decide when to stop~\cite{9176839}, problems like the arbitrariness of discretising continuous parameters and redundancy/bias in scenario generation limit their credibility~\cite{amersbach2019defining}.

\textit{ii) What is the residual safety risk after extensive scenario-based testing?} Even after testing a vast number of scenarios (say achieved $100\%$ coverage of some criteria), the residual risk (measured by, e.g., the probability of crash per random scenario/mile) remains unknown. This uncertainty arises because coverage criteria typically measure which parts of the input space have been exercised but do not translate directly into an estimate of overall system safety delivered
~\cite{hamlet1990partition,frankl1998evaluating} (a well-known problem in software engineering). 

\textit{iii) Is scenario-based testing always more effective at improving safety of AVs?} Scenario-based testing ``smartly'' selects conditions that are hypothesised to stress the AV, thus potentially detecting more bugs than mile-based testing. However, ``more bugs does not necessarily imply less reliable''~\cite{bev2007icse}. Consider an extreme example: if scenario-based testing uncovers only ``5,000-year bugs''\footnote{E. Adams studied fault occurrence rates in large IBM software systems and found that over 60\% were ``5,000-year bugs'', meaning each fault manifests as failures, on average, once every 5,000 years of operation~\cite{adams1980minimizing}.}, fixing them may have little improvement on AVs' operational safety, compared to fixing fewer ``5-year bugs'' discovered through mile-based testing. Essentially, \textit{not all bugs contribute equally to operational safety risk}. 
Scenario-based testing, e.g., using optimisation algorithms to achieve high coverage, often overlooks this fact in its objectives, while mile-based testing naturally detects those bugs likely to manifest themselves in operation.

\textit{iv) How does the fidelity\footnote{While specific definitions of fidelity vary across studies, in general, fidelity refers to how closely synthetic data replicate real-world conditions, thereby supporting reliable simulation-based testing of AVs.} of simulated scenarios affect safety estimates?} Given that scenario-based testing focuses on high-risk conditions, synthetic data generated by simulation is typically employed to execute test scenarios. However, the Sim2Real gap introduces an additional layer of uncertainty, implying that safety claims based on synthetic scenario testing evidence may not fully capture real-world performance \cite{iso_scenario_2022}. Despite great efforts having been made to bridge the gap~\cite{stocco2022mind,hu2023simulation,cheng2024instance,yan2023learning}, there remains a lack of rigorous methods to quantify simulation fidelity~\cite{cheng2024instance} and translate varying fidelity levels into \textit{measurable} safety risk.

The aforementioned questions are not independent but rather highly interrelated.
Without rigorous answers to them, the usefulness of scenario-based testing in \textit{formal safety arguments for regulatory justification} may be limited. While there is no simple solution, we believe that \textit{statistical analysis and probabilistic modelling can provide a solid theoretical foundation} for addressing these challenges.

In this paper, we begin by reviewing the state-of-the-art in statistical modelling for AV safety assessment. We observe that most existing work focuses on mile-based testing, while statistical foundations for scenario-based testing remain largely under-explored, despite its growing prominence in industry practice. Furthermore, we notice that many of the open questions in scenario-based testing, e.g., effectiveness in debugging and residual risk estimation, \textit{closely parallel longstanding challenges studied in the traditional software testing community}. We present an initial model, as a proof-of-concept, to demonstrate the potential of applying statistical approaches to rigorously reason about scenario-based safety evidence, and highlight the potential for AV testing to benefit from software testing, with existing models potentially being reusable. Our goal is to lay a foundation that future work can build upon to develop rigorous safety arguments for AVs.

In summary, the main contributions of this paper are:
\begin{itemize}[leftmargin=*]
    \item A survey on statistical modelling for AV safety assessment, and a comparison study with software testing to map shared research questions and identify reusable solutions.
    \item Proof-of-concept models using scenario-based testing to assess AV safety and quantify the reduced operational risk.
    \item An example measure of \textit{Risk Estimation Fidelity} of synthetic data, with formal definitions and discussion of its possible use towards simulation-based safety arguments.
\end{itemize}

\section{AV Testing vs Software Testing}
\subsection{AV Testing}
The main idea behind scenario-based testing for AVs is to generate \textit{relevant scenarios} intentionally in simulations or on proving grounds, rather than waiting until they \textit{randomly} occur in distance-based testing. Relevant scenarios are all scenarios that can occur within the ODD and are challenging for the AV and, therefore, might cause failures\footnote{\label{footnote_failure}In this paper, ``failure'' is used as a general term in the context of AV safety. Depending on the assessor’s focus in specific models, it may refer to events such as a crash, fatality, near-miss, or human-initiated disengagement.}.
\begin{definition}[ODD~\cite{bsi_scenario_2023}]
An ODD is a set of operating conditions under which a given driving automation system or feature thereof is specifically designed to function.
\end{definition}
\begin{definition}[Scenario~\cite{ulbrich2015defining}]
A scenario describes the temporal development between several scenes in a sequence of scenes, where a scene is a snapshot of the environment, including the scenery and dynamic elements, and all actors’ and observers’ self-representations and the relationships among those entities.
\end{definition}

\begin{definition}[Scenario abstraction levels~\cite{menzel2018scenarios}]
Scenarios are commonly organised into four levels of abstraction:
1) Functional scenarios use natural language to describe high-level situations.
2) Abstract scenarios formalise natural language descriptions into machine-readable formats, often supported by ontologies.
3) Logical scenarios define parametrised versions of abstract scenarios, specifying variable ranges (e.g., speed, position, weather).
4) Concrete scenarios assign specific values to each parameter, representing fully specified test cases derived from logical scenarios.
\end{definition}
Note, such levels of abstraction can also be viewed as levels of partitioning in the parameter space that describes scenarios.

\begin{definition}[Mile-based testing]
Mile-based testing evaluates AV safety by accumulating real/simulated miles within the ODD, aiming to sample scenarios as they naturally occur. Safety is inferred statistically by analysing failure rates (e.g., crashes, disengagements) per unit of operation. It assumes an unbiased estimate and exposure to diverse scenarios (thus it may offers some benefits similar to scenario-based testing).
\end{definition}

Similar to~\cite{amersbach2019defining}, the following remark attempts to link the two types of AV testing, with a relaxation of rigour in details, e.g., the overlapping of scenarios on driven miles.

\begin{remark}[Mile-based testing vs scenario-based testing]
\label{remark_mile_vs_scenario}
Mile-based testing can be viewed as a form of scenario-based testing, where driven miles can be modelled as a stochastic sequence of concrete scenarios sampled from the ODD according to an ``operational scenario distribution''.
\end{remark}

\subsection{Software Testing}

There is a substantial body of work in the software engineering community on safety and reliability modelling based on testing evidence. While software testing methods can be classified from various perspectives, e.g., functional vs non-functional, black-box vs white-box, and static vs dynamic, we focus on the following two dimensions that facilitate mapping concepts from AV testing to software testing.

Broadly, software testing falls into two categories: 
\textit{directed} testing and \textit{operational} testing; they are also often called \textit{debug} testing and \textit{acceptance} testing \cite{frankl1998evaluating, cotroneo2015relai} based on their typical purposes. The former aims to find and fix defects, while the latter assesses existing quality to build confidence for deployment. The key difference is whether test cases are sampled from the operational profile (OP). Debug testing typically ignores the OP, whereas acceptance testing depends on it~\cite{frankl1998evaluating}. However, research shows operational testing may outperform directed testing in bug detection, and some acceptance tests rely solely on directed methods. We show this using our proof-of-concept models later.

\begin{definition}[OP~\cite{musa1993op}]
An OP is a probability distribution over the input space 
of a software system, representing the relative frequencies with which different input sequences expected occur during actual operation.
\end{definition}

From the perspective of ``\textit{how we generate tests}'' in the input parameter space, software testing can be broadly categorised into \textit{partition-based testing} and \textit{random testing} \cite{ntafos1998random,368132}.

\begin{definition}[Partition-based testing]
Partition-based testing partitions the input domain of a software system into distinct ``subdomains/partition''.
Test cases are then selected from each subdomain, ensuring that at least one is selected from each subdomain.
\end{definition}
This approach aims to maximise testing efficiency by ``smartly'' partitioning the input space and selecting representative test cases from each partition, thereby reducing the total number of test cases while maintaining effective coverage.
\begin{definition}[Random testing]
Random testing generates test cases randomly from the input domain, according to a predefined probability distribution, e.g., the OP. 
\end{definition}

\subsection{Mapping Concepts Between AV and Software Testing}

Let us consider the parameter space used to describe scenarios (within a given ODD) as analogous to the input parameter space in software testing. Then we may conclude:
\begin{remark}[ODD vs OP]
The ODD can be regarded as an OP without explicitly defining the distribution over it. If an ``operational scenario distribution'' can be defined over the ODD, conceptually, it aligns with the OP in software testing.  
\end{remark}
\begin{remark}[Scenario-based testing vs partition-based testing]
\label{remark_scenario_as_partition}
Scenario-based testing can be viewed as a form of partition-based testing, where logical scenarios partition the ODD, and representative concrete scenarios are picked for testing in each partition to ensure coverage. Both aim to save tests by assuming that with clever partitioning, failures will cluster in certain 
subdomains, allowing testers to focus effort where it matters most.
\end{remark}
\begin{remark}[Mile-based testing is random testing]
\label{remark_mile_as_random}
Continuing Remark~\ref{remark_mile_vs_scenario}, since each concrete scenario arises randomly from an operational scenario distribution (i.e., the OP), mile-based testing resembles randomly selected test cases (with replacement) across the scenario parameter space.
\end{remark}
\begin{remark}[Align AV testing with testing purpose]
Given the motivation, assumptions, and design of partition-based and random-based testing, they naturally align with debug testing and acceptance testing (where an OP is used), respectively. Thus, following Remarks \ref{remark_scenario_as_partition} and \ref{remark_mile_as_random}, scenario-based testing aligns more closely with the goals of debugging, whereas mile-based testing is better suited for acceptance testing.
\end{remark}
That said, there are exceptional cases that we learnt from studies of software testing, e.g.,~\cite{frankl1998evaluating,bev2007icse,littlewood_validation_1993,hamlet1990partition,cotroneo2015relai,ntafos1998random,368132}: 1) Partition/scenario-based testing is often thought to be good for debugging, but there is no formal proof and it may be less effective than operational testing
under certain, not especially unusual, conditions. 2) Partition/scenario-based testing can still be used for estimating reliability (i.e., residual safety risk), if distributional information is explicitly used within, and across, logical scenario partitions. 3) Random/mile-based testing can still discover bugs, more or less efficiently, in terms of improving safety, than partition/scenario testing. All of these points are context-dependent and cannot be answered universally. This highlights \textit{the need for a theoretical model with statistical foundation capable of analysing different AV testing methods across conditions}.

\section{Statistical Modelling for AV Safety}
\label{sec_stat_model}

There are, indeed, many existing works on AV safety applying statistical reasoning to estimate failure rates. However, they typically focus on specific abstract or logical scenarios (e.g.,~\cite{zhao2016accelerated,zhao2017accelerated,ren2025intelligent,feng2021intelligent,ding2020learning}), rather than modelling across the entire ODD using an operational distribution---which is the focus of our work. We also exclude papers that rely solely on descriptive statistics for AV safety, and instead focus on those that employ statistical inference where an underlying stochastic failure process is assumed. To the best of our knowledge, the selected publications are summarised in Table~\ref{tab_summary}, from which the following key insights can be drawn.

\begin{table*}[th]
\centering
\caption{Summary of AV safety studies employing statistical inference with assumed stochastic failure processes across operational domains.
}
\label{tab_summary}
\begin{tabular*}{1 \linewidth}{@{\extracolsep{\fill}}|m{1cm}|m{1cm}|m{1.3cm}|m{2.5cm}|m{3cm}|m{5.5cm}|}
\hline
\textbf{Paper} & \textbf{Metric} & \textbf{Stoc. Proc.} & \textbf{Estimator} & \textbf{Data form and Source} & \textbf{Main Message} \\
\hline
\hline

\cite{kalra_driving_2016} & \emph{pfm} & Bernoulli & \makecell[l]{Reliability model\\ from \cite{practical_oconnor_2011}} & \makecell[l]{Failure-free miles,\\ illustrative data} & \makecell[l]{Extensive mileage requirements.\\Necessity for alternative methods.\\Adaptive regulatory frameworks.} \\
\hline

\cite{kalra_driving_2016} & \emph{fpm} & Poisson & \makecell[l]{Significance test} & \makecell[l]{Miles with failure,\\ illustrative data} & \makecell[l]{Extensive mileage requirements.\\Necessity for alternative methods.\\Adaptive regulatory frameworks.} \\
\hline


\cite{zhao2019assessing,zhao_assessing_2020} & \emph{pfm} & Bernoulli & CBI & \makecell[l]{Failure-free \& rare failure\\ illustrative data} & \makecell[l]{In addition to \cite{kalra_driving_2016}, prior knowledge may bring \\ advantages, while avoiding optimistic biases.} \\
 \hline

\cite{zhao2019assessing} & \emph{dpm} & Poisson & SRGM & \makecell[l]{Waymo’s disengagement\\ data (up to 2019)} & \makecell[l]{Sensitive to model used.\\CBI preferred for high-reliability claims.} \\
\hline

\cite{Min_recurrent_2022} & \emph{dpm} & NHPP & \makecell[l]{Monotonic splines\\ (non-parametric model)} & \makecell[l]{CDMV's disengagement\\ data} & \makecell[l]{AVs are becoming more reliable by \\decreasing disengagement trends \\Statistical tools useful for monitoring.} \\
\hline

\cite{Zheng04032025} & \emph{dpm} & NHPP & \makecell[l]{Bayesian inference\\ with SRGM} & \makecell[l]{CDMV's disengagement\\ data} & \makecell[l]{Providing a statistically grounded structured\\ approach to AV reliability test design
} \\
\hline

\cite{Zheng04032025} & \emph{dpm} & HPP & \makecell[l]{Bayesian inference\\ without SRGM} & \makecell[l]{CDMV's disengagement\\ data} & \makecell[l]{Providing a statistically grounded structured\\ approach to AV reliability test design
} \\

\hline

\end{tabular*}

\vspace{-5mm}
\end{table*}

\begin{remark}[Metrics]
\label{remark_metric}
   We can model the process of failure occurrence as occurring in a sequence of discrete trials (a \emph{discrete time} model), so that the parameter of interest is a probability  of failure per trip, scenario, etc., or in response to continuous exposure (e.g. number of miles driven or time driving), and in this \emph{continuous time} model the parameter of interest is a \emph{rate} of occurrence.
   The choice between discrete and continuous models determines the underlying stochastic process (e.g., Bernoulli for probabilities or Poisson for rates) and informs whether a discrete or continuous model is more appropriate.
    The metric also guides how data should be collected in practice, and whether the data realistically supports the assumptions of the chosen stochastic process. 
       
\end{remark}
Several other works~\cite{littlewood_reliability_2020, Zhao_2021_DSN, bishop_bootstrapping_2022, Aghazadeh_fault_freeness, Aghazadeh_impact_2024} study the probability of failure per demand (\emph{pfd}), a general metric applicable across domains such as AVs and nuclear power. In the context of AV safety, a ``demand'' can be abstracted as a trip, a mission, or a scenario, as long as the modelling assumptions can be justified at the defined ``demand'' level, e.g., AV trips are independently and identically distributed (i.i.d.).



\begin{remark}[Stochastic process]
\label{remark_stochastic_proc}
For mathematical convenience, common stochastic processes are used, e.g., Bernoulli (for \textit{pfm}) or Poisson (for \textit{dpm}). However, as discussed in the papers listed in Table. \ref{tab_summary}, those  stochastic processes impose assumptions---e.g., a constant failure probability/rate per mile and independence between miles/events---that may not always hold true in real-world applications due to factors like software updating of the AV or changing environments. That said, they can still offer a useful approximation in practice. 
\end{remark}

After defining the stochastic process, likelihood functions of the failure events of interest can be derived. Then either Bayesian or frequentist estimators can be applied.  
As shown in Table~\ref{tab_summary}, statistical significance testing, CBI (Conservative Bayesian Inference), and SRGM (Software Reliability Growth Models) are the most commonly used estimation methods. 
CBI uses evidence-based \textit{partial prior knowledge} about the metric (e.g., \textit{pfm}) to determine a set of prior distributions\footnote{Classical Bayesian inference requires a fully specified prior distribution, which is often difficult to justify in practice.}.
This set is then updated using Bayesian inference based on statistical operational evidence, and the most conservative posteriors can be derived by optimisation.
SRGMs, originating from software reliability engineering, are meant to extrapolate observed trends as a software product becomes more reliable as bugs are found and corrected. The models themselves may for instance represent the failure process as a non-Homogeneous Poisson Process (NHPP).

\begin{remark}[Estimator]
\label{remark_estimator}
Various standards require ``frequentist'' estimates of confidence in bounds on the metric of interest. These may yield less confidence than desired, and incorrect estimates of the metric when too few data are available. Bayesian methods on the other hand can take into account prior knowledge about the AV, sometimes giving sufficient confidence despite limited data; but the reliability of the estimates depend on proper representation of the actual prior knowledge \cite{Aghazadeh_impact_2024}.
\end{remark}
\begin{remark}[Estimator]
SRGMs help track and predict failures over time, but they cannot be trusted a priori. They are especially problematic for highly reliable software and, even when generally accurate for a system, can occasionally produce seriously incorrect predictions. 
\end{remark}

\section{Proof-of-Concept Models}

\subsection{The Metric, Stochastic Failure Processes, and Estimators}
As stated in Remark~\ref{remark_metric}, the choice of metric is the first step in statistical inference.
As a first attempt, we align with existing metrics and define the \textit{probability of failure per randomly selected scenario} (drawn from the operational scenario distribution over the ODD), abbreviated as \textit{pfs}:
\begin{definition}[\textit{pfs}]
Let $x\in D$ be a concrete scenario in the scenario parameter space $D$ of the ODD, and $Op: D \rightarrow [0,1]$ is the operational scenario distribution.
\begin{equation}
    \textit{pfs}:=\int_{x \in D}I_{\{\text{x causes a failure}\}}(x)Op(x)dx = \theta
\end{equation}
where $I_{\{S\}}$ is an indicator function---it is equal to 1 when $S$ is true and 0 otherwise. We then denote \textit{pfs} as $\theta$.
\end{definition}
Similar to the ``frequentist'' interpretation of \textit{pfd} in~\cite{zhao_modeling_2017}, \textit{pfs} can be regarded as a \textit{limiting relative frequency of concrete scenarios for which the AV fails in an infinite sequence of independently selected concrete scenarios from the OP}. Note, as stated in footnote \ref{footnote_failure}, the ``failure'' in \textit{pfs} is a general term that can be replaced by any safety event of the assessor's interest, e.g., crash and near-miss.

\subsubsection{Scenario-based testing conducted in a mile-based/random testing way}
The Bernoulli process normally is a good approximation of the stochastic failure process for this case \cite{kalra_driving_2016,zhao2019assessing}, thus the likelihood of seeing $k$ failed scenarios in a sequence of $t$ random test (concrete) scenarios is:
\begin{equation}
    Pr(k, t \mid \theta)= \theta^k(1-\theta)^{t-k}.
\end{equation}
likelihood function determines which statistical inference methods can be used to estimate $\theta$. These include those methods listed in Table \ref{tab_summary} and the choice depends on the observed data (e.g., whether they include failures), prior knowledge available, and desired posterior estimation forms (e.g., mean or confidence bounds of \textit{pfs})
\begin{equation}
\label{eq_post_mean_est}
   \hat{\theta} = \mathbb{E}[\theta \mid k,t]=\frac{\int_{0}^{1} \theta \cdot \theta^k(1-\theta)^{t-k} f(\theta) d\theta}{\int_{0}^{1} \theta^k(1-\theta)^{t-k} f(\theta) d\theta}
\end{equation}
where $f(\theta)$ is the prior distribution of \textit{pfs}. Thanks to the set of CBI models summarised earlier, a complete $f$ is not required; rather, partial knowledge of $f$ would suffice for the reasoning. 

\subsubsection{Scenario-based testing conducted in a partition-based testing way} Let the scenario parameter space $D$ be partitioned into subdomains $D_1,\dots,D_n$, i.e., $n$ logical scenarios. Then, by the law of total probability
\begin{equation}
    \theta=\sum_{i=1}^{n}\theta_iOp_i, \text{ where } Op_i:=\sum_{x\in D_i}Op(x)
\end{equation}
Thus, to estimate $\theta$, we need to know the \textit{conditional} \textit{pfs} for each logical scenario $\theta_i$ and the OP. Normally, inside a given logical scenario $D_i$, a new and (thought-to-be) more efficient distribution is used to generate concrete test scenarios (e.g., by an optimisation algorithm) rather than the (local) OP. Usually, the estimator using \textit{importance sampling} can be used where the proposal distribution is either known or implicitly approximated~\cite{zhao2016accelerated,zhao2017accelerated,ren2025intelligent,feng2021intelligent}. 

\subsection{Under Which Conditions Will Scenario-based Testing Deliver Better Safety than Mile-based Testing?}
We continue to investigate the above question by retrofitting the model from \cite{frankl1998evaluating} for AV testing. To model repair behaviour after detecting failures, \cite{frankl1998evaluating} introduces the concept of ``failure region ($F$)''---a set of inputs in $D$ that cause failure: 
 a fix removes a specific failure region and does not introduce new failure regions\footnote{The original paper~\cite{frankl1998evaluating} discusses the validity of some assumptions and model extensions to relax them, e.g., ``partial fix''. For simplicity, we omit these discussions here.}. While~\cite{frankl1998evaluating} considers the more general case of multiple (disjoint) failure regions, we only present the simplified \textit{single failure region} case to better illustrate the conceptual mapping between AV and software testing, and the potential for reusing existing research insights.

Consider an AV under test whose \textit{pfs}, $\theta=q$, is due to a single failure region. After mile-based testing that included $t$ random concrete scenarios (cf. Remarks \ref{remark_mile_vs_scenario} \& \ref{remark_mile_as_random}), the failure region will have been detected, and fixed, or not. By the formula of total probability, the resulting expected \textit{pfs}  
is as below, 
where the ``$0$'' reflects the fact that after fixing, no bugs remain.

\vspace{-4mm}
\begin{equation}
\label{eq_exp_pfs_op_testing}
    \mathbb{E}[\theta]=0\cdot Pr(\theta=0)+q \cdot Pr(\theta=q)=q(1-q)^t
\end{equation}

Now consider scenario-based testing and let the scenario parameter space $D$ be partitioned into subdomains $D_1,\dots,D_n$, i.e., $n$ logical scenarios. According to some specific scenario-based testing methods, $t_i$ test cases are generated independently for each $D_i$. Let $d_i$ be the bug detection rate for logical scenario $D_i$, i.e., the probability that a concrete scenario generated for $D_i$ 
reveals a failure, then the probabilities of seeing at least 1 failure and no failures after scenario-based testing are, respectively: 
\begin{equation}
    1-\prod_{i=1}^{n}(1-d_i)^{t_i}, \quad \prod_{i=1}^{n}(1-d_i)^{t_i}.
\end{equation}
For the former case, after fixing the failure detected, $\theta=0$; for the latter case, the AV remains as it is with $\theta=q$, thus the expected \textit{pfs} after scenario-based testing and fixing is:
\begin{equation}
\label{eq_exp_pfs_scenario_testing}
    \mathbb{E}[\theta]=0\cdot Pr(\theta=0)+q \cdot Pr(\theta=q)=q \prod_{i=1}^{n}(1-d_i)^{t_i}
\end{equation}
For a fair comparison, we normally take $\sum_{i=1}^{n} t_i=t$. Then a comprehensive analytical study can be conducted to compare Eq.~\eqref{eq_exp_pfs_op_testing} and Eq.~\eqref{eq_exp_pfs_scenario_testing}. For simplicity, we only present examples to show that there is \textit{no universal conclusion} on the superiority of scenario-based testing over mile-based testing.

Consider two possible cases: \textit{1)} The single failure region~$F$ is spread across all possible $D_i$, e.g., a misperception failure that may happen in many logical scenarios; \textit{2)} $F$ is strictly contained by a single $D_i$, e.g., a rare failure can only happen in a specific logical scenario. We show examples for each case.
\begin{example}
When the failure region is uniformly ``spread out'' over all the subdomains, implying that the chance of finding a failure is the same in all subdomains (a constant detection rate $d_i=\bar{d}$). Then the case reduces to scenario-based testing without subdomains, and Eq.~\eqref{eq_exp_pfs_scenario_testing} becomes:
\begin{equation}
\label{eq_exp_pfs_scenario_testing_constant_d}
    \mathbb{E}[\theta]=q (1-\bar{d})^{t}
\end{equation}
Comparing Eq.~\eqref{eq_exp_pfs_scenario_testing_constant_d} to Eq.~\eqref{eq_exp_pfs_op_testing} shows that scenario-based testing is superior if and only if $ \bar{d}>q$. 
\end{example}

\begin{example}
Consider the failure region is a strict subset of a single subdomain $k$, $F \subset D_k$, and with some points in $D_k$ not being failure points, $D_k \not\subset F$, and no other subdomains overlapping with $F$: $ F \cap D_i = \emptyset, \forall i \neq k$.
Additionally, we assume that within $D_k$, both scenario-based and mile-based testing (on average) have an equal probability of encountering~$F$. That is, the detection probability $d_k$ is just the fraction of the operational distribution scenarios in $D_k$ that encounter~$F$:
$d_k = \frac{\sum_{x \in F} Op(x)}{\sum_{x \in D_k} Op(x)},$
For scenario-based testing, assume test budgets are distributed equally across $n$ subdomains, so \mbox{$t_k=\frac{t}{n}$}, where~$t$ is the total number of tests.
\begin{itemize}[leftmargin=*]
    \item \textbf{Situations when scenario-based testing is superior:}
     This case occurs when the failure-prone subdomain is relatively rare in real operation, resulting in fewer tests under mile-based testing compared to scenario-based testing.
    Mathematically, this is expressed as
$t\sum_{x \in D_k} Op(x) \ll t_k$. After substituting $t_k=\frac{t}{n}$, we will have: 
\begin{equation}
\label{eq_probability_scenario_within_Dk}
        \sum_{x \in D_k} Op(x) \ll \frac{1}{n}
\end{equation}
Then, the expected pfs for scenario-based testing is:
{\small
\begin{align}
\mathbb{E}[\theta]=&q \prod_{i=1}^n (1 - d_i)^{t_i} = q (1 - d_k)^{t_k} \label{eq_scenario_testing_Dk_1} \\
= & q\! \left(1\! -\! \frac{\sum_{x \in F} Op(x)}{\sum_{x \in D_k}\! Op(x)}\right)^{\frac{t}{n}} \!\!\!\!
<  q\! \left(1\! -\! \frac{\sum_{x \in F} Op(x)}{1/n}\right)^{\frac{t}{n}} \label{eq_scenario_testing_Dk_3} \\
\approx & q \left(1 - t \sum_{x \in F} Op(x)\right) \label{eq_scenario_testing_Dk_4} 
\approx q (1 - q)^t 
\end{align}}\normalsize
where the last term $q (1 - q)^t$  is the \textit{pfs}. 
The two approximations in
Eq.~\eqref{eq_scenario_testing_Dk_4} require that $d_k$ and $q$ are small, using $(1 + x)^y
\approx1 + yx$ for small $x$.

Intuitively, this happens because the scenario-based testing allocates more tests to the only ``failure prone'' logical scenario (subdomain $D_k$), compared to mile-based testing. 


\item \textbf{Situations when mile-based testing is superior:}
A similar argument leads to the opposite conclusion when mile-based testing heavily samples $D_k$ (many mile-based test scenarios fall within $D_k$). If there are many other subdomains, scenario-based testing ends up ``wasting'' most of its effort on them (still assuming $t_k = t/n $), rather than the ``failure prone'' subdomain. Mathematically, mile-based testing provides more tests, $ t \sum_{x \in D_k} Op(x) \gg t_k$, substituting $t_k = \frac{t}{n}$, we have:
\begin{equation}
\label{eq_probability_mile_within_Dk}
   \sum_{x \in D_k} Op(x) \gg \frac{1}{n}  
\end{equation}
Substituting Eq. \eqref{eq_probability_mile_within_Dk} in place of Eq. \eqref{eq_probability_scenario_within_Dk} in Eq. \eqref{eq_scenario_testing_Dk_3} reverses the inequality, showing that scenario-based testing results in a lower expected \textit{pfs} than mile-based testing.

\end{itemize}
\end{example}

\subsection{Modelling Fidelity of Synthetic Data}

In the scope of using synthetic data for testing AV, substantial work has been done~\cite{stocco2022mind,hu2023simulation,cheng2024instance,yan2023learning}. Those studies define fidelity in different ways, e.g.,: 1) distance between individual real and synthetic data points, 2) distance between (safety-relevant) outputs when processing real vs synthetic inputs \cite{cheng2024instance}, and 3) statistical distribution similarity of real vs synthetic datasets \cite{yan2023learning}. In this section, from the perspective of sameness in doing statistical inference for safety risk, we introduce \textit{Risk Estimation Fidelity} (REF) as:
\begin{definition}[$(\epsilon,\alpha)$-REF]
Let $\hat{\theta^r}$ and $\hat{\theta^s}$ be the \textit{pfs} estimates by an estimator\footnote{The same estimator should be used for both real and synthetic data to ensure the estimation differences reflect data bias, not estimator bias.}(e.g.,~Eq.~\eqref{eq_post_mean_est}) using real and synthetic data, respectively, then the simulator achieves $(\epsilon,\alpha)$-REF if
\begin{equation}
\label{eq_REF}
    Pr(|\hat{\theta^s}-\hat{\theta^r}|\leq \epsilon)\geq 1-\alpha
\end{equation}
where $(\epsilon,\alpha)$ is a specified tuple of two small positive constants.
\end{definition} 
Eq. \eqref{eq_REF} is a classical confidence statement\footnote{This is not a Bayesian probability distribution of the actual error, but rather a frequentist confidence bound: if you repeated the process of testing and deriving the estimator, many times, the estimator would fall within the error range ($-\epsilon, +\epsilon$) around the true value with at least $1-\alpha$ probability.
}: with confidence at least $1-\alpha$, the synthetic-data estimate deviates\footnote{In real safety cases, relative error often matters more than absolute error.} from the real-data estimate by no more than~$\epsilon$.
Noted that, even if the estimates using real and synthetic data are close, they may still detect very different failures. Thus, our goal is \textit{not to replace} existing fidelity definitions and studies like \cite{cheng2024instance}.

Let $\theta^r$ and $\theta^s$ be the (unknown) ground truth \textit{pfs} of the AV deployed in real-world and simulators, respectively. The estimation difference can be decomposed into:
\begin{align}
|\hat{\theta}^s - \hat{\theta}^r|
&= |\underbrace{(\hat{\theta}^s - \theta^s)}_{\text{sim. sampling variance}}
\!\!+ \underbrace{(\theta^s - \theta^r)}_{\text{sim. bias } \Delta}
+ \!\!\!\underbrace{(\theta^r - \hat{\theta}^r)}_{\text{real-w. sampling variance}}| \nonumber
\end{align}
The first and third terms representing sampling uncertainty can be reduced given more real and synthetic data collected. The second term $\Delta$ captures the simulator's \textit{inherent quality}, which is irreducible unless the simulator is reconfigured to be a better one.
By understanding those three uncertainties, we propose a REF use case with steps of:
\begin{enumerate}[leftmargin=*]
    \item \textbf{Collect real-world data} Test a fixed and limited number of scenarios using real-world data. Estimate \(\hat{\theta}^r\).
    
    \item \textbf{Generate synthetic data} Use the current simulator configuration to run a number of scenario tests. Estimate \(\hat{\theta}^s\).
    
    \item \textbf{Certify REF} Empirically check if the simulator satisfies \((\epsilon,\alpha)\)-REF for predefined \(\epsilon\) and confidence level \(1 - \alpha\).
    
    \item \textbf{If REF certification fails, increase synthetic data} to reduce the variance in \(\hat{\theta}^s\). Recompute REF. Repeat until: REF is satisfied, or additional samples yield no improvement.
    
    \item \textbf{If REF certification still fails, reconfigure simulator} to improve its accuracy, and hopefully reduce the gap between real and synthetic performance (i.e., reduce bias \(\Delta = \theta^s - \theta^r\)). Repeat from Step 2.
    
    \item \textbf{If REF remains unsatisfied, quantify fidelity limit} by determining the \emph{smallest} \(\epsilon\) for which REF is certified. The new $(\epsilon,\alpha)$ reflects the best REF we can estimate with the available data. If new $\epsilon$ is large, this may indicate bias in our initial limited real-world sample, or persisting problems with the simulator.
    \item \textbf{Scale-up simulation testing with certified simulators}.
    
    To use simulated results to assess the AV, we still need evidence that the simulator matches real operation, not just the limited sample we have. This requires prior evidence validating the simulator's accuracy, as we explain later
    
    \item \textbf{Monitor REF} after the AV deployment when more real-world data is collected. Recertify REF if it is violated.
\end{enumerate}

We note that REF relies on a limited initial real-world sample. 
This means that a \textit{pfs} estimate from this sample may differ from the true \textit{pfs}.
In particular, with very small \textit{pfs}, such inaccuracy will be the norm, as a small sample will generally include zero failures.
Within such a ``biased'' sample, REF may ``certify'' a simulator that happens to be affected by similar inaccuracy.
Failing to pass a REF criterion is good indication that the simulator is inaccurate or the sample ``biased'', and thus aids decisions at steps 4 and 5 above.
To prove instead that the simulator is faithful to the real world, evidence of the appropriateness of physical models and quality of development need to be combined with the statistical observations (as we mentioned at step 7): a Bayesian argument that is the goal of future work.
 Step 8 with more real-world data will further verify whether the certified-REF remains valid for the actual AVs---not just for the initial limited sample.


\begin{example}
\label{exp_ref}
We illustrate the REF workflow:
\begin{enumerate}[leftmargin=*]
    \item Step 1: Run \(t_r = 500\) on-road scenarios sampled from the OP, observing \(k_r = 17\) near-miss events (defined as the ``failure'' of our interest in this example).  Using the Maximum Likelihood Estimator (MLE):
$
  \hat\theta^r = \frac{17}{500} = 0.034.
$
\item Step 2: In simulation, run \(t_s = 2000\) comparable scenarios with \(k_s = 45\) near-misses.  By MLE,
$
  \hat\theta^s = \frac{45}{2000}=0.0225. 
$
\item Step 3: Presuming we first set error tolerance \(\epsilon = 0.02\) and confidence level \(1-\alpha = 0.95\).
For large $t_r$ and $t_s$, by the Central Limit Theorem, the two sample mean $\hat\theta^r$ and $\hat\theta^s$ follow Gaussian distributions with
\[
  \mathrm{Var}(\hat\theta^r)
  \approx \frac{\hat\theta^r(1-\hat\theta^r)}{t_r}
  = \frac{0.034 \cdot 0.966}{500}
  = 6.57\times10^{-5},
\]
\[
  \mathrm{Var}(\hat\theta^s)
  \approx \frac{\hat\theta^s(1-\hat\theta^s)}{t_s}
  = \frac{0.0225 \cdot 0.9775}{2000}
  = 1.1 \times10^{-5}.
\]
Thus the difference \(\hat{\Delta} = \hat\theta^s - \hat\theta^r\) is also a Gaussian\footnote{The sum of two Gaussian random variables is also a Gaussian one.} with
\[
  \mu_{\hat{\Delta}} = -0.0115,\quad
  \sigma_{\hat{\Delta}} 
    = \sqrt{ \mathrm{Var}(\hat\theta^s)+ \mathrm{Var}(\hat\theta^r)}
    \approx 0.0088.
\]
\begin{align}
    \Pr\bigl(|\hat{\Delta}|\le 0.02\bigr)
  &=
  \Phi\!\Bigl(\frac{0.02 - \mu_{\hat{\Delta}}}{\sigma_{\hat{\Delta}}}\Bigr)
  - \Phi\!\Bigl(\frac{-0.02 -
  \mu_{\hat{\Delta}}}{\sigma_{\hat{\Delta}}}\Bigr) \nonumber
  \\
  &\approx 0.83 < 0.95, \nonumber
\end{align}
so the simulator fails \((0.02,0.05)\)-REF with the samples.
  \item Step 4: Doubling \(t_s\) to 4000 with \(k_s=102\) yields \(\Pr(|\hat{\Delta}|\le0.02)\approx 0.91<0.95\); \((0.02,0.05)\)-REF still uncertified.
\item Step 5: Improving the simulator, collect new data of \(t_s=2000\) and \(k_s=58\), and may get
\[ \hat{\Delta} =  -0.005, \,\,
  \sigma_{\hat{\Delta}}\approx 0.0089,\, \,
  \Pr(|\hat{\Delta}|\le0.02)>0.95,
\]
so \textbf{\((0.02,0.05)\)-REF is certified}. 
\item Step 6: Skipped, as the ideal, predefined $(\epsilon,\alpha)$ tuple for REF is certified with the available data. 
\item Step 7: Scale up the $(0.02,0.05)$-REF simulator testing with $t_s = 50{,}000$ and $k_s = 1415$, we compute $\hat{\theta}^s = 0.0283$ and its standard error as $\sigma_{\hat{\theta}^s} = 0.000741$. The $95\%$ confidence interval (CI) for $\theta^s$ is:
$$\hat{\theta}^s \pm 1.96 \times \sigma_{\hat{\theta}^s} = [0.02685,\ 0.02975].$$


\end{enumerate}
\end{example}


How to incorporate prior knowledge about the simulator’s accuracy in Step 7, and how to integrate additional real-world evidence from Step 8 for estimating $\theta^r$, using both simulated and real-world data, requires proper Bayesian modelling.
In practice, trust in a simulator will generally depend on quality of the physical models implemented, of the development, etc; leading to a Bayesian argument for confirming or weakening the initial trust through simulation results. 
More work is required also to address limitations of our example, e.g. that if failures are rare the sample means can no longer be approximated by normal distributions.
Our intent here is just to demonstrate that formal probabilistic treatment of simulation accuracy is not just desirable, but likely feasible, rather than provide a comprehensive formal treatment.

A future direction is to define the mentioned Bayesian version of REF, where beliefs about the actual $\theta^r$ and $\theta^s$ are modeled, and being able to reason under rare failures. Once such a Bayesian REF is certified (e.g., with 95\% probability that $\theta^r$ is no worse than $\theta^s$), existing models for ``No Worse Than'' reasoning~\cite{zhao_assessing_2020, Zhao_2021_DSN, littlewood_reliability_2020, Aghazadeh_impact_2024, Aghazadeh_fault_freeness} can be applied to support safety claims based on both real and simulated data.

\section{Discussion and Conclusion}

The four high-level questions posed in the introduction are tightly interrelated. The stopping rule for scenario-based testing should depend on the level of residual risk that assessors are willing to accept. Estimating this residual risk, in turn, must account for the extra uncertainty introduced by synthetic data fidelity. Minimising such risk requires a clear understanding of the debug effectiveness of the test method within the available budget. So, 
\textit{centring around the residual safety risk}, 
these questions can be jointly addressed by the proof-of-concept models proposed in this paper, with more careful designs to tackle practical challenges in future.

\paragraph{The metric} While we introduce \textit{pfs} in our proof-of-concept models for scenario testing evidence, is \textit{pfs} truly the best metric? Or should we consider higher abstraction level of ``per trip'' or ``per mission''? What constitutes an acceptable \textit{pfs} in AVs? Can we derive such thresholds from human-driven incident statistics at ``per scenario'' level? Answer to those question depends on how data are collected in practice for the specific AV under study, considering, e.g., if scenarios are i.i.d. \cite{salako2023unnecessity} and if \textit{pfs} is constant for the data collected.

\paragraph{Dynamic OP} How can we accurately estimate the OP, i.e., the operational scenario distribution over the given ODD? More importantly, how can we cope with the dynamic OP over time (due to, e.g., new roads and roadworks)? While this is an established problem \cite{adams1996total,bishop_deriving_2017,pietrantuono2020reliability}, models require retrofit to cope with new challenges imposed by AV, e.g., scalability.

\paragraph{Asymmetric fidelity} In practice, false positives (simulated failures that never occur on-road) may be tolerable, whereas false negatives (missed hazards in simulation) are catastrophic. Our REF definition can be improved to cope with such practical consideration to ease the REF certification.

\paragraph{Ultra-high reliability} For AVs, demonstrating ultra-high reliability---a level unachievable through testing alone---is a core challenge~\cite{littlewood_validation_1993}. Bayesian methods (e.g., CBI) enable combining test data with prior knowledge, but identifying suitable AV-specific priors and formally translating them into beliefs about \textit{pfs} remains an open, case-dependent problem.

We caution that applying statistics in safety-critical domains requires great care: statistical inference, when used improperly, can be dangerously misleading. Thus, a solid statistical foundation is essential.
Our goal is not to replace existing methods to AV safety, but to offer a complementary perspective: one that is driven by reasoning about the \textit{residual operational risk}. By mapping challenges in AV testing to known problems in software reliability, we aim to bridge methodological gaps and enable more rigorous safety claims for AVs.


\bibliographystyle{IEEEtran}
\bibliography{ref}

\end{document}